# DoReMe: **D**ual **o**ptimization of **Re**construction and **M**otion **e**stimation for self-navigated free-breathing abdominal MRI using Gaussian Splatting

Majd Helo[1,2], Dominik Nickel[2], and Thomas Kuestner[1]

1 Medical Image and Data Analysis (MIDAS.lab), Department of Diagnostic and Interventional Radiology, University of Tuebingen, Tuebingen, Germany.

2 Research & Clinical Translation, Magnetic Resonance, Siemens Healthineers AG, Erlangen, Germany.

# Abstract

Purpose:

Free-breathing abdominal MRI is limited by respiratory motion, which introduces image artifacts and k-space inconsistencies. Existing reconstruction approaches often depend on pre-estimated motion fields or external respiratory surrogate signals. We propose a coupled reconstruction-motion estimation algorithm for self-navigated free-breathing acquisition, in which motion is represented by compact Gaussian primitives and jointly refined with the image estimates during iterative reconstruction.

Methods:

An iterative reconstruction that alternates between data consistency updates and motion modelling was proposed. Respiratory motion was represented using Gaussian primitives. A Gaussian splatting was integrated into the reconstruction algorithm. The method was evaluated on self-navigated free-breathing Cartesian gradient-echo (GRE) acquisition in four healthy volunteers, including retrospective scan shortening experiments, and was compared with regularized compressed sensing (CS) reconstruction.

Results:

Respiratory motion states were reconstructed in approximately 1.5 min. Compared with CS, the proposed method improved visual image quality, particularly in coronal liver views, and better preserved anatomical structures in retrospectively shortened acquisitions. These findings provide proof of concept for the proposed method.

Conclusion:

The proposed coupled reconstruction-motion estimation algorithm improved visual image quality compared with CS alone. Gaussian primitives provided a compact and sufficiently flexible representation of nonrigid respiratory motion with practical runtime, supporting further scanner integration and validation in larger cohorts.

# 1. Introduction

Free-breathing abdominal MRI remains limited by respiratory motion, because k-space is sampled over time while the anatomy moves continuously. Combining data from different respiratory positions without compensation introduces inconsistencies that appear as ghosting, blurring, and loss of spatial fidelity [1-3]. Breath-holding, gating, triggering, and navigator-based acceptance windows can reduce these artifacts [4-7], but often prolong or destabilize scan time, reject useful data, and remain sensitive to irregular breathing [8-11].

Accelerated MRI methods have become central to dynamic and free-breathing imaging. Parallel imaging and compressed sensing (CS) reduce sampling requirements by exploiting coil sensitivity encoding and transform sparsity [12-16]. However, acceleration alone does not remove respiratory inconsistency and can degrade the intermediate images used for motion estimation [17,11,18].

Motion-compensated reconstruction addresses this problem by incorporating motion information into image formation. Several strategies have been proposed to account for respiratory motion during reconstruction, including formulations that explicitly model motion within the imaging operator [19], coupled reconstruction–motion estimation algorithms with motion signals obtained using external physiological motion sensors [20], nonrigid motion-compensated CS reconstructions for free-breathing MRI [21,22,17,23-26] and retrospective correction based on estimated deformation fields [27-29].

Nevertheless, many motion-resolved reconstruction pipelines remain sequential: motion fields are estimated before the final reconstruction and are then treated as fixed [11,17,24,30]. Although coupled reconstruction-motion estimation algorithms like GRICS enable iterative refinement of image content and motion [20], they require motion-related surrogate information, which may be derived from external physiological sensors, navigator measurements, or other respiratory monitoring approaches. Other coupled reconstruction-motion estimation approaches instead use self-navigated acquisitions to iteratively refine both image content and motion directly from the acquired imaging data [31-34]. However, efficient representation of complex nonrigid motion within such reconstruction loops remains a key challenge, particularly when dense deformation fields must be repeatedly estimated and evaluated [35-39], motivating the development of more compact motion parameterizations. Deep learning methods can provide an alternative with rapid inference, but they generally require large training datasets and may not generalize reliably across anatomies, scanners, or acquisition protocols [40,41,18,42-46]. Implicit neural representations provide a case-specific adjustment of the motion to the patient and at the same time reduce the number of needed motion parameters. However, they are less directly interpretable and often still require iterative and computationally long optimizations [47,48,40].
Gaussian primitives offer a compact and explicit alternative. Three-dimensional Gaussian splatting represents spatial structure using anisotropic Gaussian kernels and supports efficient differentiable rendering [49]. Related registration methods have shown that Gaussian representations can enable coarse-to-fine alignment and image-guided refinement [50]. In medical image registration, GaussianDIR represents dense deformation fields between a fixed and a moving image using sparse adaptive Gaussian primitives with local K-nearest-neighbor blending [40]. While this provides a compact and interpretable alternative to dense voxel-wise motion fields, the method is formulated as pairwise image registration, which optimizes a separate deformation for each image pair.

A recent study on time-resolved volumetric MRI represents a reference-frame MR image and low-rank motion-basis components using 3D Gaussian primitives, while a dual-path MLP/CNN motion encoder estimates temporal motion-basis coefficients from k-space-derived signals. The patient-specific model is learned from a pre-treatment acquisition and subsequently applied to newly acquired online k-space data for real-time volumetric motion estimation and image generation [51]. While this approach provides an elegant solution for real-time volumetric imaging, it requires the reference image and a patient-specific low-rank motion model to be learned from a separate pre-treatment acquisition before application to subsequently acquired data.

In the present work, we propose a Gaussian representation that extends from pairwise to a groupwise motion model. Building on this representation, we propose a joint reconstruction-motion estimation algorithm for motion-resolved free-breathing MRI. Rather than reparameterizing the images themselves, Gaussian primitives are used specifically to represent respiratory deformation, while the motion-resolved images retain their conventional representation and are refined jointly with the motion model. The respiratory deformation is represented using a compact set of locally blended Gaussian primitives. The Gaussian centers, covariance parameters, and blending weights are shared across all respiratory states, while state-specific displacement vectors describe the motion associated with each primitive. This provides a compact, spatially coupled representation of respiratory deformation across the complete motion cycle. The algorithm alternates between data-consistent reconstruction updates and Gaussian splatting, allowing the motion model and image estimates to be refined iteratively. We explored the feasibility of the proposed approach in 4 free-breathing scans of healthy subjects.

# 2. Materials and Methods

The proposed algorithm for joint reconstruction and motion estimation is tailored to binned data, such as motion states, for which a regularized reconstruction aims to enforce similarity between neighboring bins and for which a group-wise motion estimation aims to align the bins. A central aspect is to use the same expression as image regularization in the reconstruction and as similarity measure in the elastic image registration. This unifies the task into a single, simultaneous optimization for images and displacement vector fields. The reconstruction part then resembles an iterative reconstruction while the motion modelling is based on Gaussian primitives.

Let $x = (x_1, \ldots, x_M)$ denote the images indexed by motion state $m$ and $\theta$ denote the parameterization of the motion model. Assuming known motion operators $U_m$ that warp a motion state $x_m$ to a reference state, a motion-resolved reconstruction that enforces similarity between neighboring motion states can be given by the optimization

$$\hat{x} = \underset{x,\theta}{\arg\min} \left[ \underbrace{\frac{1}{2} \sum_{m=1}^{M} \| A_m F S x_m - y_m \|_2^2}_{\text{Data Consistency (DC)}} + \underbrace{\frac{\lambda}{M} \sum_{m=2}^{M} \| U_m(x_m; \theta) - U_{m-1}(x_{m-1}; \theta) \|_1}_{\text{Total Variation (TV)}} + \underbrace{\rho\, R(\theta)}_{\text{Motion Constraint (MC)}} \right]. \quad [1]$$

Here the first term enforces data consistency (DC) with the acquired data $y_m$ using the sampling operators $A_m$ for the respective motion states, the Fourier transformation $F$ and the coil sensitivity operator $S$. The second term enforces similarity between neighboring motion states in the spirit of total variation with a regularization factor $\lambda$ and the third term enforces smoothness constraints on the estimated motion. The second and third term form the group-wise registration subproblem, while the first term describes the reconstruction subproblem.

The estimation of motion parameters is generally a non-convex problem and therefore depends on a suitable parameterization and initialization. In our joint reconstruction, we use an alternating optimization strategy that switches between updates of the images with fixed motion parameters and updates of motion parameters with fixed images. While the motion-estimation problem is still non-convex, this approach does not guarantee a global optimum but will converge to a local minimum.

An overview of the proposed approach is shown in Figure 1.

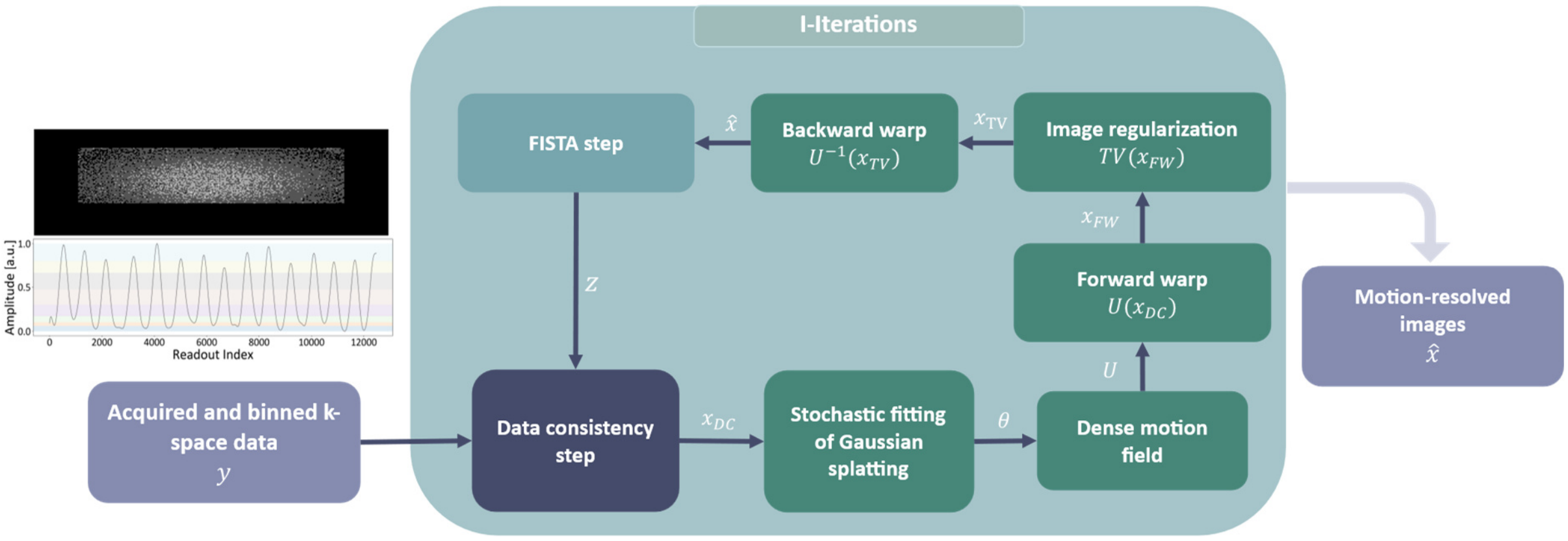

***Figure 1. Proposed Dual optimization of Reconstruction and Motion estimation (DoReMe). The acquired k-space data are first assigned to respiratory motion bins and then processed in an alternating reconstruction-motion estimation loop. Each iteration includes a DC update, fitting of the Gaussian Splatting motion model, motion-aligned regularization, and warping back to the individual motion states using the inverse deformation fields and a FISTA extrapolation step for acceleration.***

## 2.1 Motion Representation using Gaussian Splatting:

The motion operator $U_m$ is parameterized by a displacement field that in turn can be modeled using a superposition of Gaussian primitives. Indexed by $p$, these are defined as

$$G_{p,r} = \exp\left(-\frac{1}{2}\left(r - c_p\right)^T \Sigma_p^{-1} \left(r - c_p\right)\right), [2]$$

Where $r$ is the voxel location, $c_p$ is the center of the Gaussian and $\Sigma_p^{-1}$ is a symmetric positive-definite inverse covariance matrix.

The spatial influence *q* of primitive p at location *r* is given by:

$$q_{p,r} = w_p \, G_{p,r} \, , [3]$$

with weight $w_p \in \mathbb{R} > 0$. For each primitive index $p$ a k-nearest neighborhood of $K$ primitives $\bar{p} = \{p_1, \ldots, p_i, \ldots, p_K\}$ is determined and the normalized influence is defined as follows:

$$\alpha_{p,r} = \frac{\exp(q_{p,r})}{\sum_{i=1}^{K} \exp(q_{\bar{p}_i,r})} \ .[4]$$

The dense displacement field for motion state $m$ is then modeled as a superposition of Gaussian primitives, each weight with a constant displacement vector $d_{m,p}$ as

$$u_{m,r}(\theta) = \sum_{p=1}^{P} \alpha_{p,r}(\theta) \ d_{m,p} \ .[5]$$

All parameters of the multi-state Gaussian motion model are explicitly given by

$$\theta = \{c_p, \Sigma_p^{-1}, w_p, d_{m,p}\} \ .$$

The Gaussian centers $c_p$, inverse covariance matrix $\Sigma_p^{-1}$, and blending weights $w_p$ are shared across all motion states, whereas the displacement vectors $d_{m,p}$ are specific to each motion state. This parameterization provides a compact, spatially coupled representation of respiratory deformation over the complete motion cycle. Instead of a full covariance matrix $\Sigma_p^{-1}$ that also considers ellipsoidal regions of influence through diagonal and off-diagonal entries, diagonal covariance matrices were used, i.e. $\Sigma_p^{-1} = \mathrm{diag}\,(\Sigma_p^{-1})$. This reduces the number of fitted parameters while still allowing each primitive to learn different spatial extents along the image axes. Multiple overlapping primitives provide sufficient flexibility to approximate smooth respiratory motion patterns.

For fixed image estimates, the group-wise registration subproblem is defined in Eq. (1) as:

$$\hat{\theta} = \underset{\theta}{\mathrm{argmin}} \left( \frac{\lambda}{M} \sum_{m=2}^{M} \| U_m(x_m; \theta) - U_{m-1}(x_{m-1}; \theta) \|_1 + \rho \, \mathcal{R}(\theta) \right) ,[6]$$

With

$$R(\theta) = \rho_{\mathrm{smooth}} R_{\mathrm{smooth}}(\theta) + \rho_{\mathrm{fold}} R_{\mathrm{fold}}(\theta) \ ,[7]$$

where $\rho_{\mathrm{smooth}}$ and $\rho_{\mathrm{fold}}$ are the regularization factors for smoothness and folding penalties. The latter are given by:

$$R_{\mathrm{smooth}}(\theta) = \frac{1}{3MPK} \sum_{m=1}^{M} \sum_{p=1}^{P} \sum_{k=1}^{K} \| d_{m,p} - d_{m,p,k} \|_2^2 \ ,[8]$$

$$R_{\mathrm{fold}}(\theta) = \frac{1}{M\,N} \sum_{m=1}^{M} \sum_{n=1}^{N} \left( \min\left(0, J_{r+u_{m,r}(\theta)}(r_n)\right) \right)^2 \ ,[9]$$

with $K$ being the number of neighboring Gaussians, $P$ the number of Gaussians, $M$ the number of motion states, $N$ the number of voxels and $J_{r+u_{m,r}(\theta)}$ the Jacobian of the deformation fields.

The folding constraint is included to obtain invertible displacement vectors which are required for the reconstruction and computed with a fixed-point algorithm [52].

As the group-wise registration does not define a reference, an additional zero-mean displacement constraint is furthermore enforced by projecting the displacement vectors onto the zero-mean subspace across the motion states to restrict global translation:

$$d_{m,p} \;=\; d_{m,p} - \frac{1}{M}\sum_{m=1}^{M} d_{m,p} \;.[10]$$

For completeness, the motion operator (forward warp) is determined through the displacement vector through the evaluation

$$[U_m(x;\theta)]_r = \sum_{j=0}^{J-1} c_j B_j^{(3)}\left(x_{r+u_{m,r}(\theta)}\right) \;,[11]$$

Where $B_j^{(3)}$ is a third-order B-spline interpolation is applied to the right-hand side to preserve structural sharpness, and $c_j$ are the interpolation coefficients. The backward warp is obtained by inverting the motion fields using the fixed point algorithm [52].

For comparison we used a regularized CS reconstruction with TV regularization across motion states. The CS objective is defined as follows:

$$\hat{x} = \operatorname*{arg\,min}_{x}\left[\underbrace{\tfrac{1}{2}\textstyle\sum_{m=1}^{M}\|A_m F S x_m - y_m\|_2^2}_{\text{Data Consistency (DC)}} + \underbrace{\tfrac{\lambda}{M}\textstyle\sum_{m=2}^{M}\|(x_m)-(x_{m-1})\|_1}_{\text{Total Variation (TV)}}\right] \;,[12]$$

Both reconstruction methods apply TV regularization exclusively along the motion states dimension.

## 2.2 Optimization algorithm:

In case of constant displacement fields, an established algorithm to solve the minimization problem in Eq. (1) efficiently is the fast iterative shrinkage-thresholding algorithm (FISTA) [53]. This algorithm is extended by updating the motion parameters through the minimization in Eq. (6) after each FISTA-type inertial extrapolation step. This can be viewed as an inexact FISTA as the proximal operator is dependent on the iteration.

Let $H = AFS$ define the forward operator that maps the image to k-space. At iteration $i$, the gradient of the DC term is evaluated at the extrapolated image $z_i$ and gradient-descent step with the step size $\tau$ is then performed:

$$x_{DC,i} \;=\; z_i \;-\; \tau\, H^{\dagger}(H z_i - \mathrm{y})$$

The regularized image estimate $x_{i+1}$ is computed as follows:

$$x_{i+1} = U_i^{-1}\mathrm{TV}(U_i(x_{DC,i}))$$

And the FISTA update for the regularized image $x_{i+1}$ is performed using the acceleration momentum $t_i$:

$$t_{i+1} = \frac{1 + \sqrt{1 + 4t_i^2}}{2}, \qquad [13]$$

$$z_{i+1} = x_{i+1} + \frac{t_i - 1}{t_{i+1}}(x_{i+1} - x_i), [14]$$

$$\text{with: } t_0 = 1, \qquad x_0 = 0, \qquad z_0 = 0$$

The proposed algorithm converged consistently in our experiments. It can be guaranteed to converge to a local minimum when the switching to the monotone FISTA.

## 2.3 Implementation:

The processing pipeline is shown in Figure 1. The k-space data were assigned to motion states using stable amplitude binning, with readouts distributed equally across bins. Each reconstruction iteration consisted of a DC update followed by motion estimation from the current image estimates.

The GaussianDIR [40] ablation study indicated that several thousand primitives provided a favorable balance between registration accuracy and model complexity after 2000 fitting iterations. Guided by these findings, 5000 Gaussian primitives (~155000 fitting parameters: $d_{m,p}$ = 8 x 5000 x 3, $c_p$ = 5000 x 3, diagonal $\Sigma_p^{-1}$ = 5000 x 3 and $w_p$= 5000) could be used for each motion state. Within the joint reconstruction algorithm, the motion model was updated for 100 fitting iterations during each of 30 reconstruction iterations. The Gaussian centers $c_p$ were initialized randomly across the image domain. In analogy to previous coordinate-based deformable registration approaches [48,40], the image-similarity objective at each iteration was estimated from 16 000 randomly sampled voxel locations [54]. To reduce computational cost, the Gaussian model was evaluated with 64-nearest neighbors. The same stochastic sampling strategy was used to evaluate the Jacobian determinant penalty. The motion model was optimized with Adaptive Moment Estimation (Adam) [55]. At the final iteration, the Dense motion field is computed, and the images were aligned using the forward motion operator, and TV regularization was applied to the motion-compensated images. The regularized image estimates were subsequently warped back to the individual motion states using the corresponding inverse motion fields. The Gaussian primitives were refitted from the updated image estimates and initialized with the motion parameters from the previous iteration. We use a B-Spline based warping operator to avoid iterative blurring across the iterations [56].

## 2.4 Acquisition Parameters:

Data were acquired using a vendor-provided free-breathing Cartesian gradient-echo (GRE) sequence (Table 1). The sequence used a variable-density sampling pattern comprising 16 uniformly distributed masks acquired sequentially. Each mask was segmented along the partition-encoding direction into echo trains acquired after fat-suppression preparation, with echo trains ordered in ascending phase-encoding order. A self-navigation signal was extracted from a superior-inferior projection acquired before each echo train and used for k-space binning. Data were assigned to eight respiratory motion states using stable amplitude binning and reconstructed with the proposed joint reconstruction-motion estimation algorithm. Reconstructions were compared with a regularized CS reconstruction using TV regularization along the motion states dimension and with a vendor-provided breath-hold GRE acquisition obtained during inhalation with matched anatomical coverage (Table 1).

***Table 1. The acquisition protocol of the Cartesian free-breathing GRE, and the reference breath-hold measurement.***

| Acquisition Parameters | **Breath-hold GRE** | **Free-breathing GRE** |
|---|---|---|
| FOV [$mm^2$] | 309 x 380 | 309 x 400 |
| Acquired Matrix Size | 254 x 320 x 64 | 234 x 320 x 56 |
| Reconstructed Matrix Size | 338 x 320 x 96 | 312 x 320 x 88 |
| Slice Thickness [mm] | 3.0 | 3.0 |
| Orientation | Transversal | Transversal |
| Acceleration Mode/Factor | CAIPIRINHA/4 | CS/0.3 |
| Fat saturation | Spectral | Spectral |
| Slice Partial Fourier | 7/8 | 6/8 |
| Flip angle [°] | 9 | 12 |
| TR [ms] | 3.31 | 3.49 |
| TE [ms] | 1.30 | 1.32 |
| Scan Time [min:s] | 00:16 | 02:53 |

## 2.5 Retrospective Study:

Four healthy volunteers were scanned on a 3T clinical MRI system (MAGNETOM Vida, Siemens Healthineers, Forchheim, Germany) All volunteers received an informed consent discussion and provided written consent for the further use and processing of their data. MR measurements were conducted in accordance with the Declaration of Helsinki.

To assess the behavior of the motion-field estimation during optimization, reconstructed images and motion fields were visualized throughout the iterations. The estimated motion fields were overlaid as quiver plots on the corresponding image estimates.

To provide additional plausibility check of the estimated motion fields, the motion fields optimized by the proposed pipeline were compared with registration performed using Advanced Normalization Tools (ANTs)[57] on the reconstructed images. For ANTs, all motion states were registered to the middle motion state. This state was used for both methods as the reference for quantitative analysis. ANTs was configured with the Symmetric Normalization (SyN) transform. For each motion state, the Structural Similarity Index Measure (SSIM), Peak Signal-to-Noise Ratio (PSNR), Normalized Root Mean Square Error (NRMSE), High-Frequency Error Norm (HFEN), and mean absolute photometric error ($L_{1,\ Photo}$) were computed. Metrics were computed between the ANTs-/DoReMe-Registered images with the middle state as reference for both methods.  higher SSIM and PSNR and lower NRMSE, HFEN, and MAE indicate better agreement.

Additionally, we evaluated the effect of Gaussian primitive count while keeping the motion-fitting budget fixed at 100 optimization iterations per fitting step. The number of primitives was varied from 50 to 50000 across the datasets. For each configuration, the images were registered using its estimated forward deformation field. Again, we used the middle state as a reference motion state to compute the following metrics: SSIM, PSNR, NRMSE, HFEN, and photometric L1 error over the full 3D volumes and pooled across the non-reference motion states.

Retrospective undersampling experiments were performed to evaluate how both reconstruction approaches behaved under shortened acquisition conditions. For this analysis, the acquired data were retrospectively truncated by retaining only the first 50% and 25% of the readouts. Since the readouts were acquired sequentially over time, this truncation preserves the original acquisition order and mimics scans that would have been stopped earlier, rather than randomly removing k-space samples. The proposed joint reconstruction-motion estimation algorithm was then compared directly with the CS reconstruction.

The FISTA reconstruction algorithm, the Gaussian motion model and a B-Splines cubic warper were implemented in GPU using PyTorch and optimized to jointly reconstruct all respiratory motion states. Computations were performed on an NVIDIA B200 GPU with 45 GB of available video memory. Visualization of reconstructed images, motion fields, and quantitative plots was performed using Matplotlib.

# 3. Results:

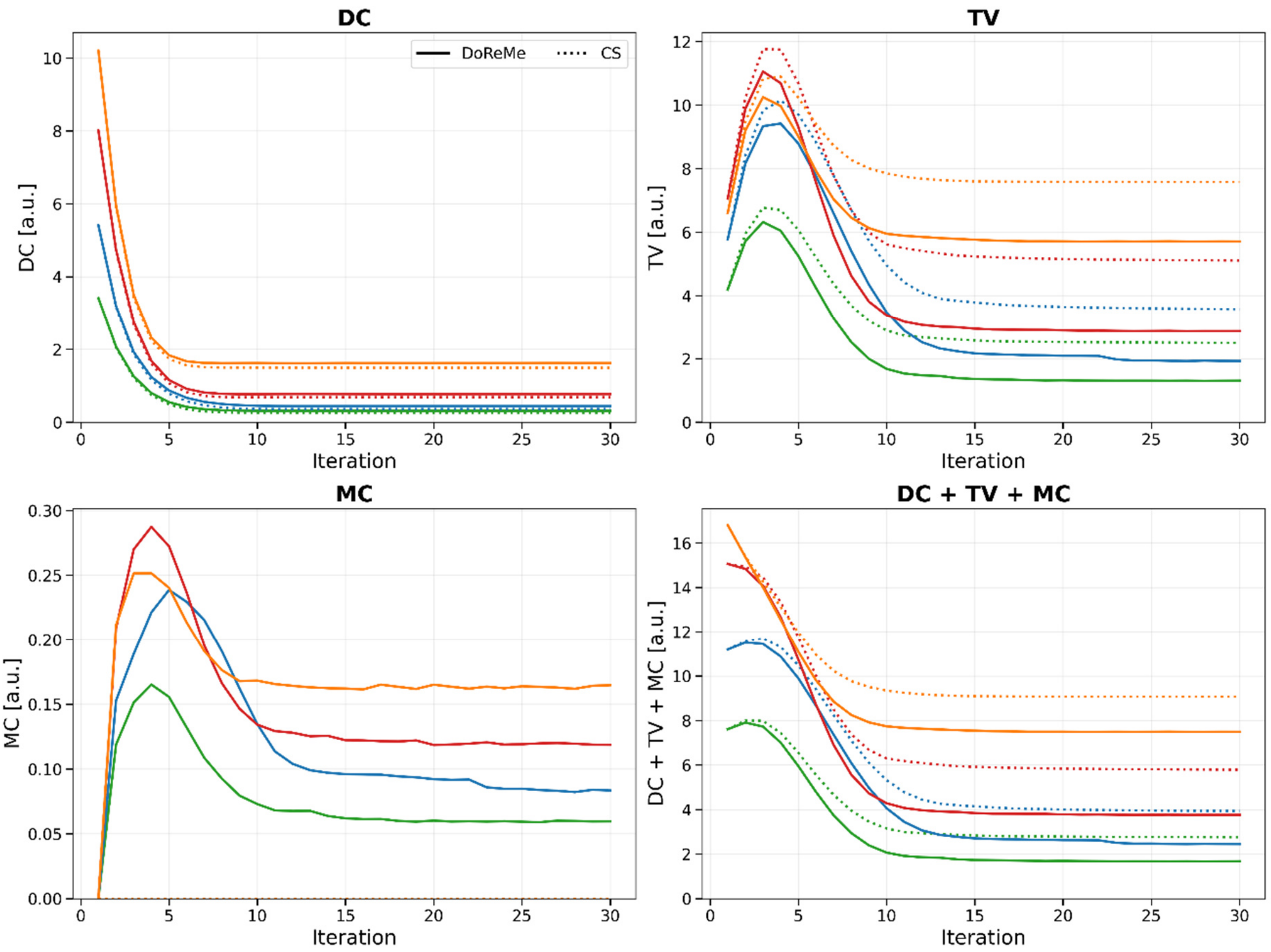

***Figure 2. Objective evolvement (Eq. 12) throughout 30 iterations. DC reconstruction error (DC), total variation (TV) regularization, motion constraint (MC), and the final objective (DC + TV + MC) are shown. Compressed sensing (CS; dotted lines) is compared with the proposed Dual optimization of Reconstruction and Motion estimation (DoReMe) algorithm (solid lines) across four volunteers, indicated by different colors. The DC error was calculated between the acquired readouts and synthetic readouts generated from the current reconstruction.***

For 30 reconstruction iterations, the FISTA reconstruction required approximately 8 s. Adding Gaussian motion-field estimation with 5000 primitives increased the runtime to approximately 45 s, and including the Jacobian-based regularization added further 45 s. Overall, the complete joint reconstruction-motion estimation algorithm required approximately 1:30 min:s, with a maximum allocated GPU memory of 14 GB for the evaluated datasets of size (Motion states x Slices x Width x Height = 8 x 88 x 320 x 312).

Figure 2 shows the evolution of the DC term, TV regularization term, motion regularization term, and total objective for CS and the proposed DoReMe algorithm across four volunteers. Compared with CS, the joint approach showed a slightly higher DC error during the iterative updates, but consistently lower TV regularization values and a lower total objective. The motion regularization term increased during the initial iterations, reached a maximum after approximately 5–7 iterations, and subsequently stabilized.

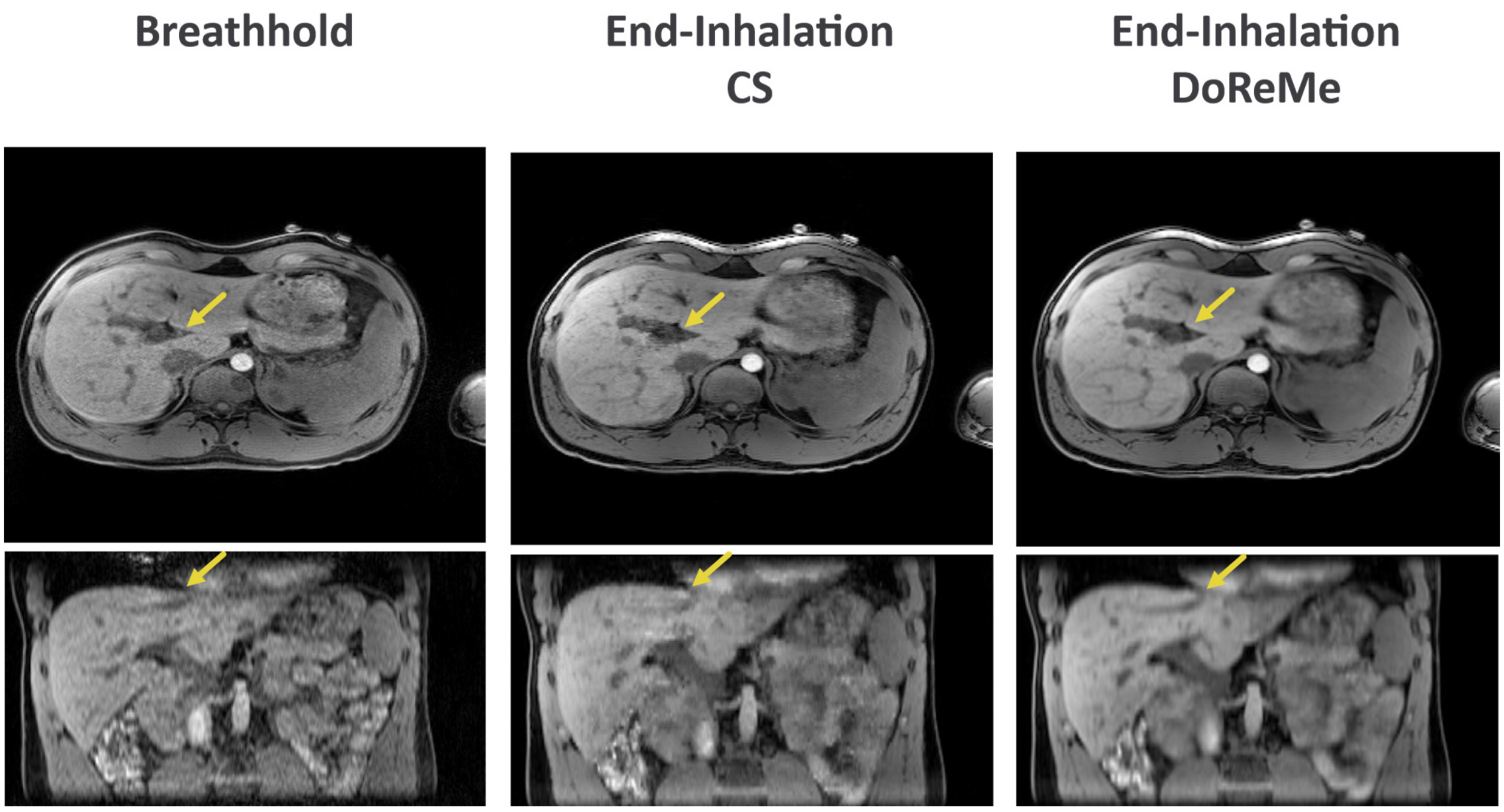

***Figure 3. Comparison of liver imaging volunteer between breath-hold (first column), free-breathing compressed sensing (CS), and the proposed Dual optimization of Reconstruction and Motion estimation (DoReMe), in the end-inhale. The top row shows breath-hold-matched transverse end-inhalation axial slices, and the bottom rows show the corresponding coronal views.***

Qualitative liver imaging showed similar trends (Figure 3). The breath-hold image was displayed alongside the end-inhalation free-breathing reconstructions to enable visual comparison across respiratory states. Transverse and coronal views were inspected. CS provided acceptable image quality. However, the proposed approach improved the depiction of hepatic boundaries and internal structures especially on coronal views.

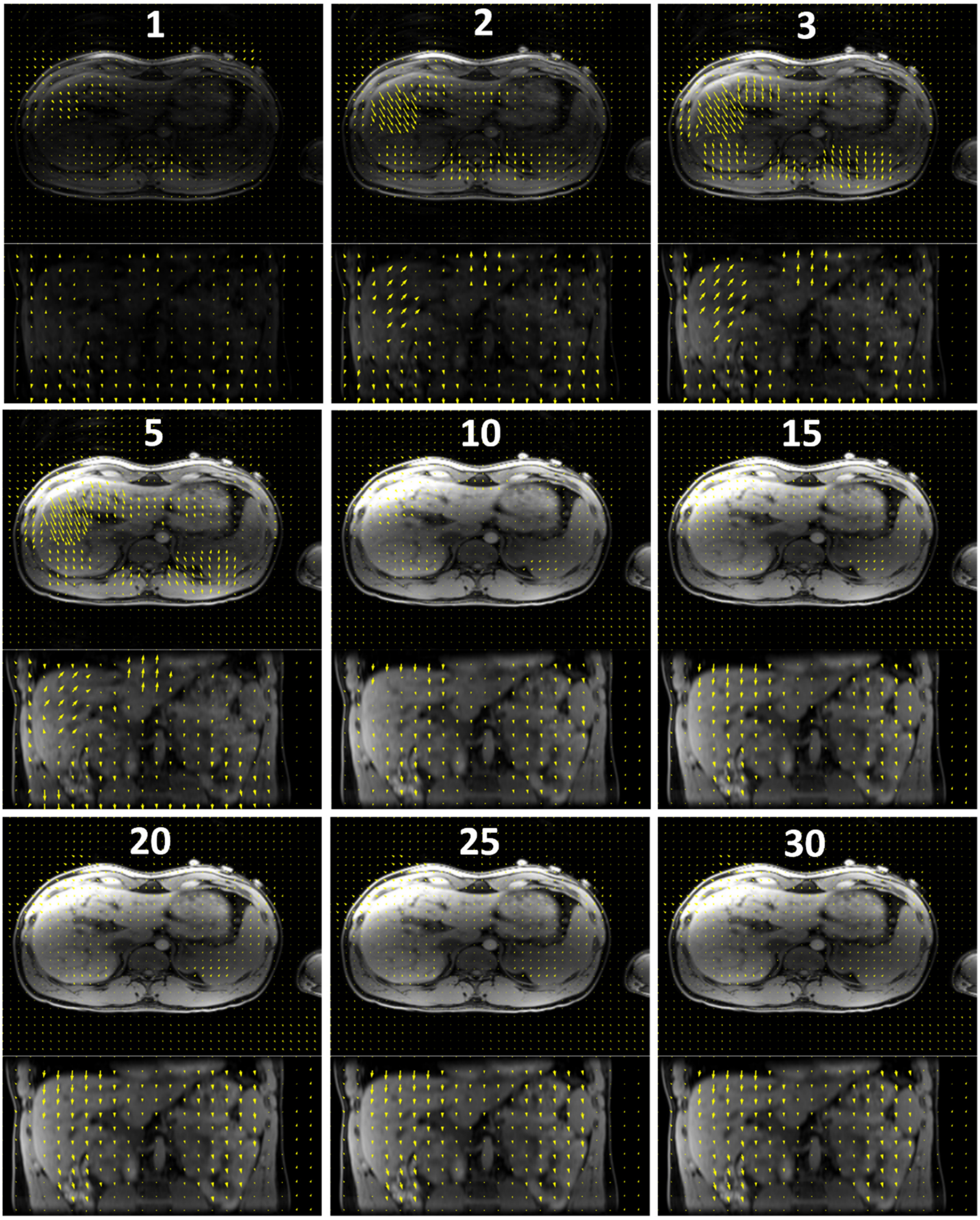

***Figure 4. Shows the evolution of the reconstructed image and estimated respiratory motion field during joint optimization, using the proposed DoReMe. Image estimates with quiver overlays are shown at iterations: 1, 2, 5, 10, 15, 25 and 30. The image and motion estimates barely changed after iteration 20.***

The evolution of the reconstructed images and estimated motion fields during optimization is illustrated in Figure 4. Early iterations showed progressive refinement of image contrast and motion-field structure, whereas later iterations produced only minor visual changes, with convergence observed around iterations 18–20. The estimated respiratory motion became spatially coherent during joint optimization and was refined together with the reconstructed image content.

***Table 2. Quantitative comparison of the proposed groupwise motion-field optimization method (DoReMe) and Advanced Normalization Tools (ANTs) registration on reconstructed images. All motion states were registered to the middle motion state, which served as the reference for ANTs. Metrics were computed after registration relative to the middle motion state. Performance was evaluated using the Structural Similarity Index Measure (SSIM), Peak Signal-to-Noise Ratio (PSNR), Normalized Root Mean Square Error (NRMSE), High-Frequency Error Norm (HFEN), and mean absolute photometric error ($L1_{photo}$). Deformation-field regularity was assessed using the percentage of voxels with a non-positive Jacobian determinant. Mean values and standard deviations are reported.***

| **Metric** | **Unregistered CS** | **ANTs** | **DoReMe (proposed)** |
|---|---|---|---|
| SSIM | 0.9576 ± 0.0249 | 0.9754 ± 0.0104 | **0.9866 ± 0.0078** |
| PSNR | 37.99 ± 3.65 | 41.62 ± 2.28 | **44.27 ± 3.09** |
| NRMSE ($\times10^{-2}$) | 1.37 ± 0.57 | 0.86 ± 0.22 | **0.65 ± 0.20** |
| HFEN ($\times10^{-1}$) | 3.88 ± 1.53 | 2.16 ± 0.93 | **1.83 ± 0.91** |
| $L_{photo}$ ($\times10^{-1}$) | 1.20 ± 0.49 | 1.14 ± 0.30 | **0.80 ± 0.31** |
| % $\{J\phi \leq 0\}$ | — | **0.0000 ± 0.0000** | 0.0143 ± 0.0192 |

The quantitative evaluation of the registration performance is reported in Table 2. Relative to unregistered CS results, both registration approaches improved inter-state alignment. DoReMe achieved the highest SSIM (0.9866 ± 0.0078) and PSNR (44.27 ± 3.09 dB), compared with (0.9754 ± 0.0104) and (41.62 ± 2.28) dB for ANTs and (0.9576 ± 0.0249) and (37.99 ± 3.65 dB) for unregistered FISTA, respectively. DoReMe also yielded the lowest NRMSE (0.65 ± 0.20 $\times 10^{-2}$), HFEN (1.83 ± 0.91 $\times 10^{-1}$), and photometric error (0.80 ± 0.31 $\times 10^{-1}$), compared with 0.86 ± 0.22 $\times 10^{-2}$, 2.16 ± 0.93 $\times 10^{-1}$, and 1.14 ± 0.30 $\times 10^{-1}$ for ANTs. ANTs produced no voxels with non-positive Jacobian determinants, whereas DoReMe showed a small fraction of 0.0143 ± 0.0192%.

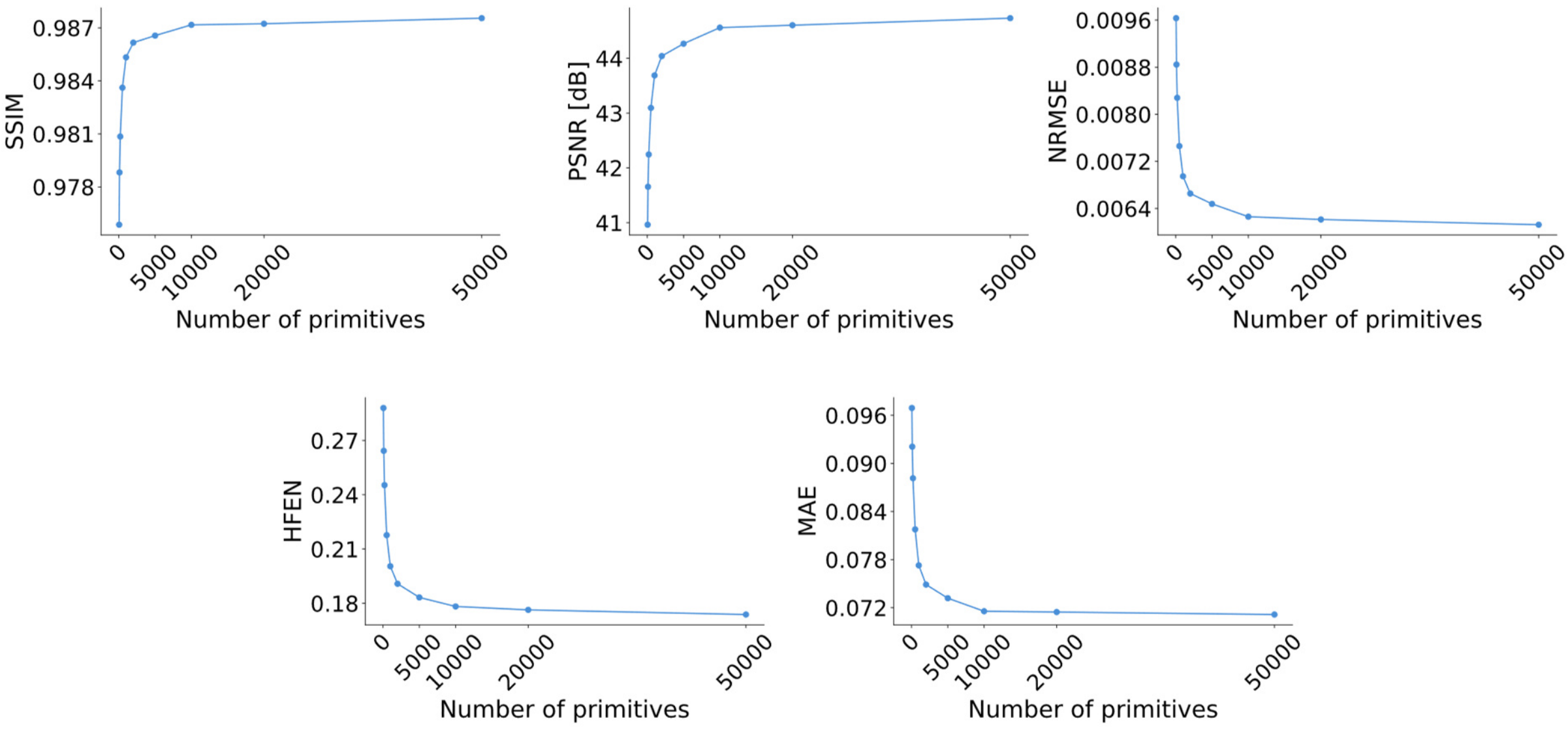

***Figure 5. Effect of Gaussian model size on motion-registration quality. Each panel shows the pooled mean metric as a function of the number of Gaussian primitives. The metrics SSIM, PSNR, NRMSE, HFEN, and mean absolute error (MAE) are computed between the registered states using the output deformation field and a reference middle state. Results are shown for motion-fitting iterations and are pooled across datasets and all non-reference respiratory states. Full-volume metrics, including background voxels, are reported.***

Figure 5 shows the effect of Gaussian model size on motion-registration quality. Metrics were computed between the registered respiratory states using the output deformation field, and the middle state as reference. The reported values represent pooled means across datasets and all non-reference states over the full image volume, including background voxels. Increasing the number of Gaussian primitives from 50 to 50000 improved all evaluated metrics: SSIM increased from 0.9759 ± 0.0144 to 0.9875 ± 0.0071, and PSNR increased from 40.96 ± 3.43 to 44.73 ± 3.02 dB, while NRMSE decreased from 0.0096 to 0.0061, HFEN from 0.2880 to 0.1738, and MAE from 0.0969 to 0.0712. The improvements became smaller beyond approximately 5000–10 000 primitives. Increasing the model size from 5000 to 50000 primitives improved SSIM by only 0.0009 and PSNR by 0.46 dB, while the mean reconstruction time increased from 170.84 to 462.60 s. Thus, 5000 primitives provided a practical compromise between registration quality and runtime. These metrics quantify agreement between registered respiratory states rather than reconstruction accuracy against an independent ground truth.

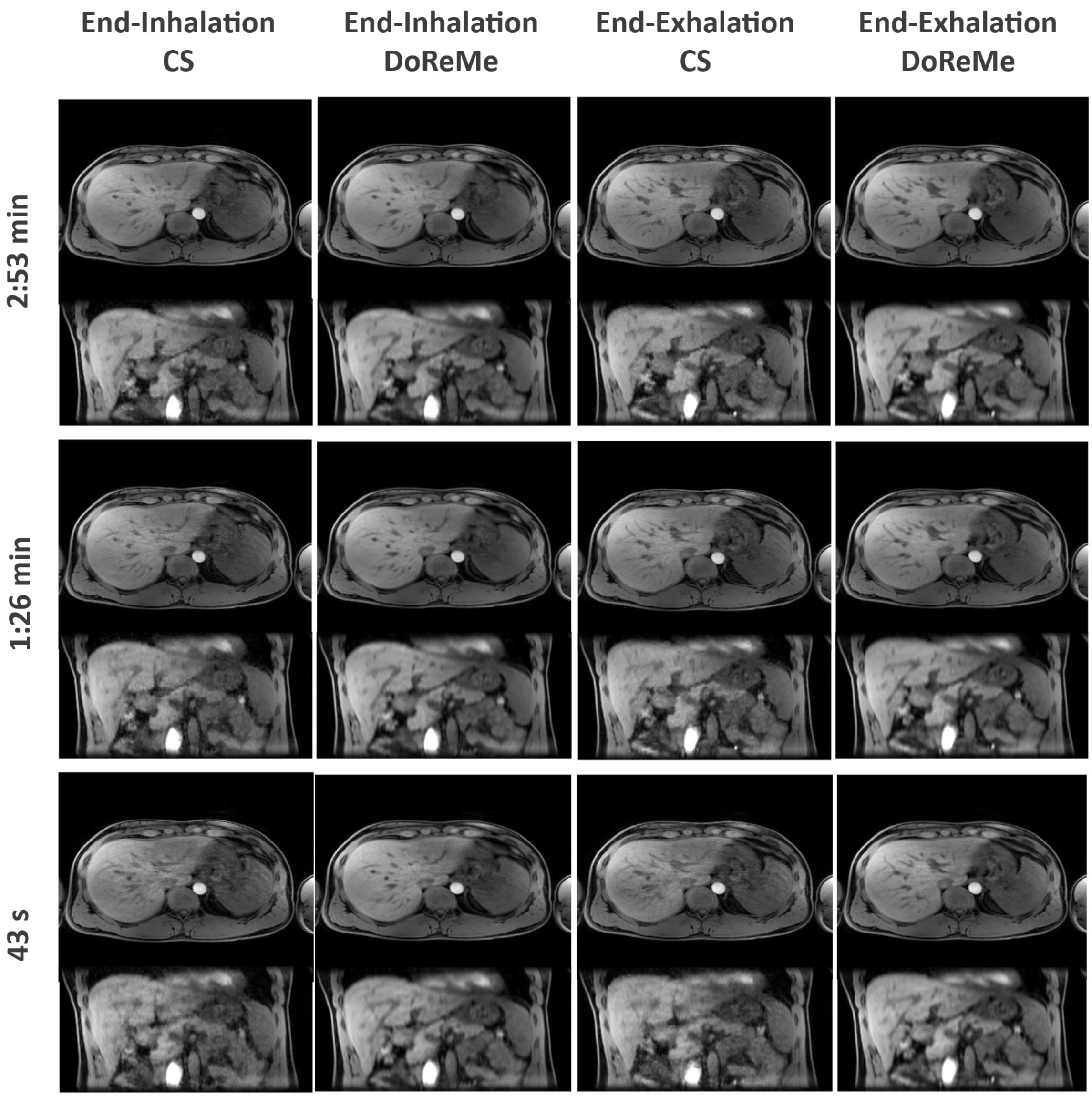

***Figure 6. Retrospective undersampling experiment showing transverse and coronal views at end-inhalation (first two columns) and end-exhalation (last two columns), for a Compressed Sensing (CS) reconstruction and the proposed DoReMe. Rows show the fully sampled reconstruction, 50% retrospective undersampling, and 75% retrospective undersampling. The proposed DoReMe preserved higher visual image quality than CS under accelerated acquisition conditions.***

Retrospective undersampling results are shown in Figure 6. End-exhalation and end-inhalation images were reconstructed using the fully sampled data as well as retrospectively shortened acquisitions retaining 50% (1:26 min) and 25% (43 s) of the acquired readouts. Across both transverse and coronal views, the proposed joint reconstruction preserved higher visual image quality than CS alone under undersampling. This indicates that the joint approach remained robust when fewer readouts were available, supporting its potential for accelerated free-breathing acquisitions.

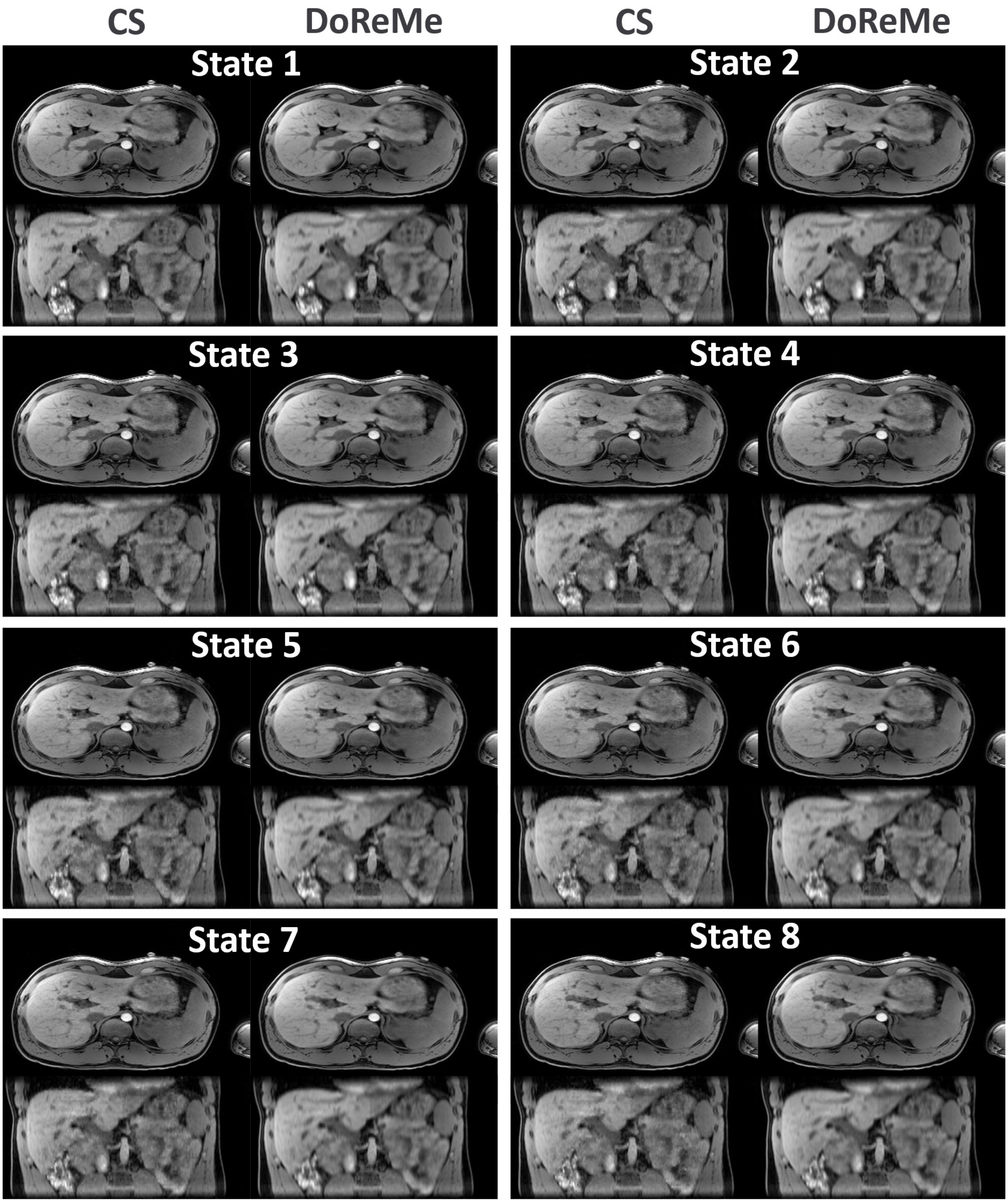

*Figure 7. Free-breathing volunteer scan reconstructed into eight respiratory motion states. CS reconstruction is shown in the left column of each state and the proposed DoReMe method in the right column, with axial and coronal views displayed. DoReMe produced more consistent anatomy and improved image quality across the respiratory states, especially in states 4–8.*

Figure 7 shows a free-breathing volunteer scan reconstructed into eight respiratory motion states. For each state, CS and the proposed DoReMe reconstruction are displayed side by side in axial and coronal views. Across the respiratory cycle, DoReMe provided more consistent anatomical appearance than CS, with improved visual image quality and clearer structural delineation. These differences were particularly evident in states 4–8, where the CS reconstructions showed stronger residual inconsistencies, whereas DoReMe maintained better anatomical continuity across the motion-resolved images.

# 4. Discussion:

This study presents a joint reconstruction-motion estimation algorithm for motion-resolved free-breathing abdominal MRI that represents respiratory deformation with compact Gaussian Splatting primitives. By iteratively refining image estimates and motion fields, the proposed approach improved visual image quality compared with CS alone. The Gaussian primitives provide an explicit, localized, and compact parameterization of nonrigid respiratory motion, avoiding independent voxel-wise deformation estimates while remaining suitable for repeated fitting during reconstruction. The GPU-accelerated implementation performed the alternating reconstruction and motion-estimation updates in approximately 1.5 minutes for the evaluated datasets.

The qualitative results suggest that the proposed method can reduce residual respiratory artifacts and improve anatomical delineation in free-breathing liver imaging. Compared with CS alone, the joint reconstruction provided a sharper depiction of hepatic boundaries and internal structures, particularly in the coronal views. These improvements indicate that iterative refinement of the deformation model can reduce inconsistencies between respiratory motion bins and make more effective use of the acquired k-space data across different breathing positions. DoReMe produced lower TV regularization values and a lower final objective than CS, despite a slightly higher DC error. This indicates that the joint method reached a different balance between data fidelity and regularization. The regularization terms changed primarily during the early iterations and subsequently converged to stable values. Overall, the lower final objective suggests that the joint framework achieved a more favorable solution according to the complete optimization criterion. This might be attributed to the lower TV error when aligning the motion states.

The evolution of the image estimates and quiver-based motion fields demonstrates that the proposed group-wise model captured smooth and coherent respiratory motion across the complete cycle. By sharing the Gaussian centers, covariance parameters, and blending weights across motion states while using state-specific displacement vectors, the model coupled the individual deformations within a common spatial representation.

Compared with ANTs, the proposed method showed good registration performance, with higher SSIM and PSNR and lower photometric error NRMSE, and HFEN. ANTs produced no non-positive Jacobian determinants, whereas the proposed method showed only a very small fraction (0.0143 ± 0.0192%), indicating slightly lower deformation regularity but largely physically plausible motion fields.

Our ablation study showed that 5000 Gaussian primitives provided sufficient flexibility to represent respiratory and liver motion while remaining compact enough for repeated optimization within the joint reconstruction loop. Although registration quality improved as the model size increased, the gains diminished beyond approximately 5000–10000 primitives: increasing the model from 5000 to 50000 primitives improved SSIM by only 0.0009 and PSNR by 0.46 dB, while increasing the mean reconstruction time from 170.84 to 462.60 s. This suggests that additional primitives mainly increase computational cost once the dominant spatial variation of the respiratory deformation has been captured. Thus, 5000 primitives represent a practical trade-off between registration quality and runtime for the evaluated datasets. However, because these metrics quantify agreement between registered respiratory states rather than accuracy against an independent ground truth, this model-size choice should be validated further in datasets with known motion.

Our findings for groupwise motion parameters are thus consistent with the ablation study of GaussianDIR for pairwise motion [40], where Li et al. showed that several thousand adaptive Gaussian primitives can approximate a dense deformation field between a moving and reference image with fewer explicitly optimized parameters than voxel-wise models. The stabilization of the displacement fields over iterations suggests that the compact Gaussian representation supports iterative motion refinement without dense independent motion parameters.

Under shortened acquisitions, CS alone became more sensitive to residual motion and undersampling artifacts, whereas the proposed approach better preserved image structure across respiratory states. This reflects the advantage of updating the deformation model together with the image estimate: as the reconstruction improves, the motion estimate can be refined, and the refined motion field can in turn improve the consistency of data acquired at different respiratory positions.

The eight-state free-breathing reconstruction further illustrates the benefit of coupling reconstruction and motion estimation across the respiratory cycle. Compared with CS, DoReMe produced more consistent anatomy and improved image quality across motion states, especially in states 4–8, where residual inconsistencies were more apparent in the CS reconstruction. This suggests that iterative refinement of the compact Gaussian motion model can improve anatomical continuity not only at selected end-inhalation or end-exhalation positions, but also across intermediate respiratory states.

Beyond the presented liver imaging experiments, the proposed algorithm may be relevant for other free-breathing applications in which respiratory motion limits image quality or anatomical reliability. One potential application is motion-resolved MRI for radiotherapy planning, where images across the breathing cycle are needed to characterize organ and target positions during treatment simulation. In this setting, jointly updating image content and respiratory motion fields could improve anatomical delineation in patients with irregular breathing or limited breath-hold capability. Another potential application entails free-breathing dynamic contrast-enhanced (DCE) MRI, where respiratory motion and contrast dynamics occur simultaneously and can compromise the consistency of motion-resolved images. Because the motion model is refined directly from the evolving reconstruction, the proposed approach may support more robust Cartesian DCE reconstruction, particularly in patients with irregular or poor breath-holding performance.

Several limitations remain present. A phantom with known motion should be acquired and used to evaluate the motion estimation model. The study was performed in a small volunteer cohort. Further validation with reader assessments are needed before clinical use. In addition, the method depends on respiratory binning, which may affect the estimated motion fields. Finally, the alternating reconstruction-registration loop is computationally more demanding than CS alone, and the current Python implementation is not yet integrated on the scanner for prospective studies.

Future work should evaluate the method in larger cohorts and in patient populations with irregular breathing. Integration on the scanner for more task-specific clinical studies would further clarify the practical benefit of the proposed joint reconstruction-motion estimation approach.

# 5. Conclusion:

A joint reconstruction-motion estimation algorithm for motion-resolved free-breathing abdominal MRI was introduced. The algorithm represents respiratory deformation using Gaussian Splatting. By alternating between data-consistent image reconstruction and refitting of the compact motion model, the proposed method avoids treating motion estimation as a fixed preprocessing step and enables iterative coupled refinement of both image content and respiratory motion fields.

In volunteer experiments, the joint approach improved image quality compared with CS alone. The method preserved higher visual image quality across respiratory states, remained robust in retrospective undersampling experiments, and improved image quality in a rapid-breathing case with pronounced motion artifacts. These findings suggest that compact Gaussian Splatting motion models can provide an effective representation for coupling nonrigid motion estimation with accelerated free-breathing MRI reconstruction.

Further validation in larger cohorts and optimization of computational performance will be required before clinical translation. Nevertheless, the proposed closed reconstruction-registration loop provides a promising direction for motion-resolved free-breathing MRI with improved robustness to respiratory motion and shortened acquisitions.

# References:

1. Wang Y, Riederer SJ, Ehman RL (1995) Respiratory motion of the heart: kinematics and the implications for the spatial resolution in coronary imaging. Magnetic resonance in medicine 33 (5):713–719.

2. Manke D, Nehrke K, Börnert P, Rösch P, Dössel O (2002) Respiratory motion in coronary magnetic resonance angiography: a comparison of different motion models. Journal of Magnetic Resonance Imaging: An Official Journal of the International Society for Magnetic Resonance in Medicine 15 (6):661–671.

3. Shechter G, Ozturk C, Resar JR, McVeigh ER (2004) Respiratory motion of the heart from free breathing coronary angiograms. IEEE transactions on medical imaging 23 (8):1046–1056.

4. Ehman RL, Felmlee JP (1989) Adaptive technique for high-definition MR imaging of moving structures. Radiology 173 (1):255–263.

5. Pipe JG (1999) Motion correction with PROPELLER MRI: application to head motion and free-breathing cardiac imaging. Magnetic Resonance in Medicine: An Official Journal of the International Society for Magnetic Resonance in Medicine 42 (5):963–969.

6. Taylor AM, Keegan J, Jhooti P, Firmin DN, Pennell DJ (1999) Calculation of a subject-specific adaptive motion-correction factor for improved real-time navigator echo-gated magnetic resonance coronary angiography. Journal of Cardiovascular Magnetic Resonance 1 (2):131–138.

7. Jhooti P, Gatehouse P, Keegan J, Bunce N, Taylor A, Firmin D (2000) Phase ordering with automatic window selection (PAWS): a novel motion-resistant technique for 3D coronary imaging. Magnetic Resonance in Medicine: An Official Journal of the International Society for Magnetic Resonance in Medicine 43 (3):470–480.

8. Stehning C, Börnert P, Nehrke K, Eggers H, Stuber M (2005) Free-breathing whole-heart coronary MRA with 3D radial SSFP and self-navigated image reconstruction. Magnetic Resonance in Medicine: An Official Journal of the International Society for Magnetic Resonance in Medicine 54 (2):476–480.

9. Henningsson M, Koken P, Stehning C, Razavi R, Prieto C, Botnar RM (2012) Whole-heart coronary MR angiography with 2D self-navigated image reconstruction. Magnetic resonance in medicine 67 (2):437–445.

10. Piccini D, Littmann A, Nielles-Vallespin S, Zenge MO (2012) Respiratory self-navigation for whole-heart bright-blood coronary MRI: methods for robust isolation and automatic segmentation of the blood pool. Magnetic resonance in medicine 68 (2):571–579.

11. Cheng JY, Zhang T, Ruangwattanapaisarn N, Alley MT, Uecker M, Pauly JM, Lustig M, Vasanawala SS (2015) Free-breathing pediatric MRI with nonrigid motion correction and acceleration. Journal of Magnetic Resonance Imaging 42 (2):407–420.

12. Sodickson DK, Manning WJ (1997) Simultaneous acquisition of spatial harmonics (SMASH): fast imaging with radiofrequency coil arrays. Magnetic resonance in medicine 38 (4):591–603.

13. Pruessmann KP, Weiger M, Scheidegger MB, Boesiger P (1999) SENSE: sensitivity encoding for fast MRI. Magnetic Resonance in Medicine: An Official Journal of the International Society for Magnetic Resonance in Medicine 42 (5):952–962.

14. Griswold MA, Jakob PM, Heidemann RM, Nittka M, Jellus V, Wang J, Kiefer B, Haase A (2002) Generalized autocalibrating partially parallel acquisitions (GRAPPA). Magnetic Resonance in Medicine: An Official Journal of the International Society for Magnetic Resonance in Medicine 47 (6):1202–1210.

15. Lustig M, Donoho D, Pauly JM (2007) Sparse MRI: The application of compressed sensing for rapid MR imaging. Magnetic Resonance in Medicine: An Official Journal of the International Society for Magnetic Resonance in Medicine 58 (6):1182–1195.
16. Uecker M, Lai P, Murphy MJ, Virtue P, Elad M, Pauly JM, Vasanawala SS, Lustig M (2014) ESPIRiT—an eigenvalue approach to autocalibrating parallel MRI: where SENSE meets GRAPPA. Magnetic resonance in medicine 71 (3):990–1001.
17. Usman M, Atkinson D, Odille F, Kolbitsch C, Vaillant G, Schaeffter T, Batchelor PG, Prieto C (2013) Motion corrected compressed sensing for free-breathing dynamic cardiac MRI. Magnetic resonance in medicine 70 (2):504–516.
18. Poddar S, Jacob M (2015) Dynamic MRI using smoothness regularization on manifolds (SToRM). IEEE transactions on medical imaging 35 (4):1106–1115.
19. Batchelor PG, Atkinson D, Irarrazaval P, Hill D, Hajnal J, Larkman D (2005) Matrix description of general motion correction applied to multishot images. Magnetic Resonance in Medicine: An Official Journal of the International Society for Magnetic Resonance in Medicine 54 (5):1273–1280.
20. Odille F, Vuissoz PA, Marie PY, Felblinger J (2008) Generalized reconstruction by inversion of coupled systems (GRICS) applied to free-breathing MRI. Magnetic Resonance in Medicine: An Official Journal of the International Society for Magnetic Resonance in Medicine 60 (1):146–157.
21. Kellman P, Chefd'hotel C, Lorenz CH, Mancini C, Arai AE, McVeigh ER (2008) Fully automatic, retrospective enhancement of real-time acquired cardiac cine MR images using image-based navigators and respiratory motion-corrected averaging. Magnetic Resonance in Medicine: An Official Journal of the International Society for Magnetic Resonance in Medicine 59 (4):771–778.
22. Hansen MS, Sørensen TS, Arai AE, Kellman P (2012) Retrospective reconstruction of high temporal resolution cine images from real-time MRI using iterative motion correction. Magnetic Resonance in Medicine 68 (3):741–750.
23. Prieto C, Doneva M, Usman M, Henningsson M, Greil G, Schaeffter T, Botnar RM (2015) Highly efficient respiratory motion compensated free-breathing coronary MRA using golden-step Cartesian acquisition. Journal of magnetic resonance imaging 41 (3):738–746.
24. Cruz G, Atkinson D, Henningsson M, Botnar RM, Prieto C (2017) Highly efficient nonrigid motion-corrected 3D whole-heart coronary vessel wall imaging. Magnetic resonance in medicine 77 (5):1894–1908.
25. Küstner T, Würslin C, Schwartz M, Martirosian P, Gatidis S, Brendle C, Seith F, Schick F, Schwenzer NF, Yang B (2017) Self-navigated 4D cartesian imaging of periodic motion in the body trunk using partial k-space compressed sensing. Magnetic resonance in medicine 78 (2):632–644.
26. Küstner T, Bustin A, Jaubert O, Hajhosseiny R, Masci PG, Neji R, Botnar R, Prieto C (2021) Fully self-gated free-running 3D Cartesian cardiac CINE with isotropic whole-heart coverage in less than 2 min. NMR in Biomedicine 34 (1):e4409.
27. White M, Hawkes D, Melbourne A, Collins D, Coolens C, Hawkins M, Leach M, Atkinson D (2009) Motion artifact correction in free-breathing abdominal MRI using overlapping partial samples to recover image deformations. Magnetic Resonance in Medicine: An Official Journal of the International Society for Magnetic Resonance in Medicine 62 (2):440–449.
28. Schmidt JF, Buehrer M, Boesiger P, Kozerke S (2011) Nonrigid retrospective respiratory motion correction in whole-heart coronary MRA. Magnetic resonance in medicine 66 (6):1541–1549.

29. Schrauben EM, Lima da Cruz G, Roy CW, Küstner T (2026) Motion mitigation techniques for abdominal and cardiac MR imaging. Journal of Magnetic Resonance Imaging 63 (4):917–937.
30. Zhao N, O'Connor D, Basarab A, Ruan D, Sheng K (2019) Motion compensated dynamic MRI reconstruction with local affine optical flow estimation. IEEE Transactions on Biomedical Engineering 66 (11):3050–3059.
31. Odille F, Uribe S, Batchelor PG, Prieto C, Schaeffter T, Atkinson D (2010) Model-based reconstruction for cardiac cine MRI without ECG or breath holding. Magnetic Resonance in Medicine 63 (5):1247–1257.
32. Zhu X, Chan M, Lustig M, Johnson KM, Larson PE (2020) Iterative motion-compensation reconstruction ultra-short TE (iMoCo UTE) for high-resolution free-breathing pulmonary MRI. Magnetic resonance in medicine 83 (4):1208–1221.
33. Zou Q, Torres LA, Fain SB, Higano NS, Bates AJ, Jacob M (2022) Dynamic imaging using motion-compensated smoothness regularization on manifolds (MoCo-SToRM). Physics in medicine & biology 67 (14):144001.
34. Lingala SG, DiBella E, Chefd'hotel C, Nadar M, Jacob M (2012) Motion compensated reconstruction for myocardial perfusion MRI. Journal of cardiovascular magnetic resonance 14 (Suppl 1):M11.
35. Beg MF, Miller MI, Trouvé A, Younes L (2005) Computing large deformation metric mappings via geodesic flows of diffeomorphisms. International journal of computer vision 61 (2):139–157.
36. Modat M, Ridgway GR, Taylor ZA, Lehmann M, Barnes J, Hawkes DJ, Fox NC, Ourselin S (2010) Fast free-form deformation using graphics processing units. Computer methods and programs in biomedicine 98 (3):278–284.
37. Vercauteren T, Pennec X, Perchant A, Ayache N (2009) Diffeomorphic demons: Efficient non-parametric image registration. NeuroImage 45 (1):S61–S72.
38. Buerger C, Clough RE, King AP, Schaeffter T, Prieto C (2012) Nonrigid motion modeling of the liver from 3-D undersampled self-gated golden-radial phase encoded MRI. IEEE transactions on medical imaging 31 (3):805–815.
39. Vishnevskiy V, Gass T, Szekely G, Tanner C, Goksel O (2016) Isotropic total variation regularization of displacements in parametric image registration. IEEE transactions on medical imaging 36 (2):385–395.
40. Li J, Liu X, Zhang F, Cao X, Buhmann JM, Zhang Y, Li X (2025) Gaussian primitives for deformable image registration. Physics and Imaging in Radiation Oncology:100821.
41. Mok TC, Chung A Fast symmetric diffeomorphic image registration with convolutional neural networks. In: Proceedings of the IEEE/CVF conference on computer vision and pattern recognition, 2020. pp 4644–4653.
42. Jena R, Sethi D, Chaudhari P, Gee JC (2024) Deep learning in medical image registration: Magic or mirage? Advances in Neural Information Processing Systems 37:108331–108353.
43. Ghoul A, Pan J, Lingg A, Kübler J, Krumm P, Hammernik K, Rueckert D, Gatidis S, Küstner T (2024) Attention-aware non-rigid image registration for accelerated MR imaging. IEEE transactions on medical imaging 43 (8):3013–3026.
44. Meng M, Feng D, Bi L, Kim J Correlation-aware coarse-to-fine mlps for deformable medical image registration. In: Proceedings of the IEEE/CVF Conference on Computer Vision and Pattern Recognition, 2024. pp 9645–9654.
45. Chen J, Frey EC, He Y, Segars WP, Li Y, Du Y (2022) Transmorph: Transformer for unsupervised medical image registration. Medical image analysis 82:102615.

46. Balakrishnan G, Zhao A, Sabuncu MR, Guttag J, Dalca AV (2019) Voxelmorph: a learning framework for deformable medical image registration. IEEE transactions on medical imaging 38 (8):1788–1800.
47. Van Harten LD, Stoker J, Išgum I (2023) Robust deformable image registration using cycle-consistent implicit representations. IEEE transactions on medical imaging 43 (2):784–793.
48. Wolterink JM, Zwienenberg JC, Brune C Implicit neural representations for deformable image registration. In: International Conference on medical imaging with deep learning, 2022. PMLR, pp 1349–1359.
49. Kerbl B, Kopanas G, Leimkühler T, Drettakis G (2023) 3d gaussian splatting for real-time radiance field rendering. ACM Trans Graph 42 (4):139:131–139:114.
50. Chang J, Xu Y, Li Y, Chen Y, Feng W, Han X Gaussreg: Fast 3d registration with gaussian splatting. In: European Conference on Computer Vision, 2024. Springer, pp 407–423.
51. Xie J, Shao H-C, Wu C, Otazo R, Deng J, Lin M-H, Chiu T, Buatti J, Iakovenko V, Zhang Y (2026) Spatiotemporal Gaussian representation-based dynamic reconstruction and motion estimation framework for time-resolved volumetric MR imaging (DREME-GSMR). ArXiv:arXiv: 2604.06482 v06481.
52. Chen M, Lu W, Chen Q, Ruchala KJ, Olivera GH (2008) A simple fixed-point approach to invert a deformation field a. Medical physics 35 (1):81–88.
53. Beck A, Teboulle M A fast iterative shrinkage-thresholding algorithm with application to wavelet-based image deblurring. In: 2009 IEEE international conference on acoustics, speech and signal processing, 2009. IEEE, pp 693–696.
54. Robbins H, Monro S (1951) A stochastic approximation method. The annals of mathematical statistics:400–407.
55. Diederik K (2014) Adam: A method for stochastic optimization. (No Title).
56. Meijering EH, Niessen WJ, Viergever MA (2001) Quantitative evaluation of convolution-based methods for medical image interpolation. Medical image analysis 5 (2):111–126.
57. Avants BB, Tustison N, Song G (2009) Advanced normalization tools (ANTS). Insight j 2 (365):1–35.